\documentclass[aps,prl,twocolumn,superscriptaddress,10pt]{revtex4-2}
\usepackage{graphicx}
\usepackage{color}
\usepackage{hyperref}
\usepackage{amsmath}
\usepackage{amssymb}
\usepackage{tabularx}
\usepackage{nicefrac}
\usepackage[defaultlines=2,all]{nowidow}

\begin{document}

\title{Excitation Spectrum and Superfluid Gap of an Ultracold Fermi Gas}

\author{Hauke Biss}
\affiliation{Institut f\"{u}r Laserphysik, Universit\"{a}t Hamburg}
\affiliation{The Hamburg Centre for Ultrafast Imaging, Universit\"{a}t Hamburg, Luruper Chaussee 149, 22761 Hamburg}

\author{Lennart Sobirey}
\affiliation{Institut f\"{u}r Laserphysik, Universit\"{a}t Hamburg}

\author{Niclas Luick}
\affiliation{Institut f\"{u}r Laserphysik, Universit\"{a}t Hamburg}
\affiliation{The Hamburg Centre for Ultrafast Imaging, Universit\"{a}t Hamburg, Luruper Chaussee 149, 22761 Hamburg}

\author{Markus Bohlen}
\affiliation{Institut f\"{u}r Laserphysik, Universit\"{a}t Hamburg}
\affiliation{The Hamburg Centre for Ultrafast Imaging, Universit\"{a}t Hamburg, Luruper Chaussee 149, 22761 Hamburg}

\author{Jami J. Kinnunen}
\affiliation{Department of Applied Physics, Aalto University School of Science, FI-00076 Aalto, Finland}

\author{Georg M. Bruun}
\affiliation{Center for Complex Quantum Systems, Department of Physics and Astronomy, Aarhus University, Ny Munkegade 120, DK-8000 Aarhus C, Denmark}
\affiliation{Shenzhen Institute for Quantum Science and Engineering and Department of Physics, Southern University of Science and Technology, Shenzhen 518055, China}

\author{Thomas Lompe}
\email{tlompe@physik.uni-hamburg.de}
\affiliation{Institut f\"{u}r Laserphysik, Universit\"{a}t Hamburg}
\affiliation{The Hamburg Centre for Ultrafast Imaging, Universit\"{a}t Hamburg, Luruper Chaussee 149, 22761 Hamburg}

\author{Henning Moritz}
\affiliation{Institut f\"{u}r Laserphysik, Universit\"{a}t Hamburg}
\affiliation{The Hamburg Centre for Ultrafast Imaging, Universit\"{a}t Hamburg, Luruper Chaussee 149, 22761 Hamburg}

\date{\today}

\begin{abstract}

Ultracold atomic gases are a powerful tool to experimentally study strongly correlated quantum many-body systems.
In particular, ultracold Fermi gases with tunable interactions have allowed to realize the famous BEC-BCS crossover from a Bose-Einstein condensate (BEC) of molecules to a Bardeen-Cooper-Schrieffer (BCS) superfluid of weakly bound Cooper pairs.    
However, large parts of the excitation spectrum of fermionic superfluids in the BEC-BCS crossover are still unexplored. 
In this work, we use Bragg spectroscopy to measure the full momentum-resolved low-energy excitation spectrum of strongly interacting ultracold Fermi gases.
This enables us to directly observe the smooth transformation from a bosonic to a fermionic superfluid that takes place in the BEC-BCS crossover.
We also use our spectra to determine the evolution of the superfluid gap and find excellent agreement with previous experiments and self-consistent T-matrix calculations both in the BEC and crossover regime.
However, towards the BCS regime a calculation that includes the effects of particle-hole correlations shows better agreement with our data.
\end{abstract}

\maketitle
Quantum many-body systems are ubiquitous in nature, but unless they are weakly interacting, their theoretical treatment can be extremely challenging.
An elegant solution to this problem was suggested by Landau, who realized that the low-energy excitation spectrum of a wide range of many-body systems can be understood in terms of particle-like excitations, which are adiabatically connected to the excitations of a non-interacting system~\cite{Landau1933}.
The residual interaction between these so-called quasiparticles in turn leads to the presence of collective modes, which have no counterpart in non-interacting system. 
Landau's quasiparticle theory has been spectacularly successful and is an indispensable tool for the  description of interacting many-body systems~\cite{Baym1991lfl}. 

\begin{figure}[t!]
	\center
	\includegraphics[width=8.6cm]{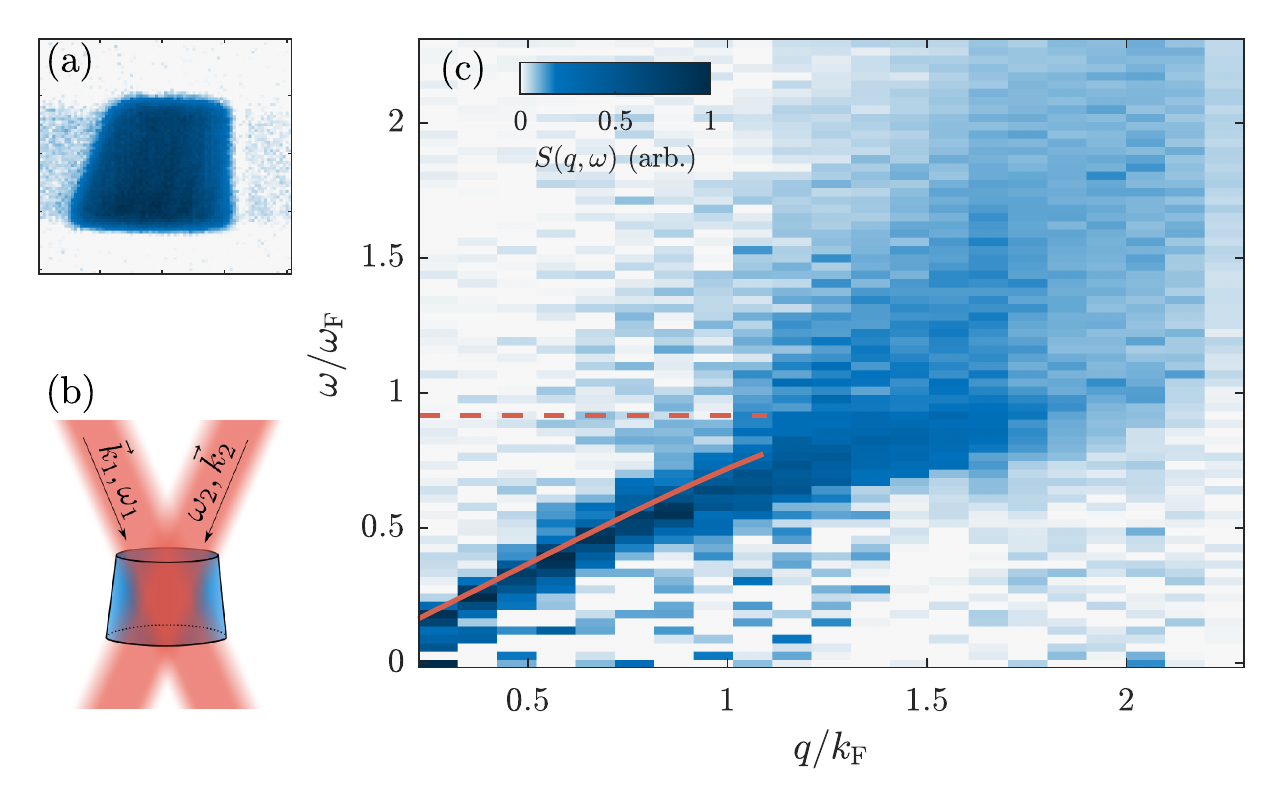}
	\caption{Measuring the excitation spectrum of an ultracold Fermi gas using Bragg spectroscopy. 
	(a) Absorption image of a homogeneous  Fermi gas trapped in an approximately cylindrical box potential.
	(b) Sketch of the experimental setup.
	Two far-detuned laser beams with frequency and wavevector $(\omega_1,\vec{k}_1)$ and $(\omega_2,\vec{k}_2)$ are used to create excitations with energy and momentum transfer $\hbar\omega=\hbar\omega_1-\hbar\omega_2$ and $\hbar q=|\hbar\vec{k}_1-\hbar\vec{k}_2|$ through a two-photon process.
	(c) Measurement of the dynamic structure factor $S(q,\omega)$ of a unitary Fermi gas.
	At low energy and momentum transfer, the Goldstone mode of the superfluid manifests itself as a linear phononic mode with a slope that corresponds to the speed of sound $v_s$.
	Pair breaking excitations occur as a broad continuum, with a clear onset at an energy corresponding to twice the pairing gap $\Delta$ of the system. 
	For comparison, the expected value of $2 \Delta$ on unitarity \cite{Haussmann2007-ph} is shown as a red dashed line, a numerical calculation of the center of the Goldstone mode is shown as a red solid line \cite{Supp}.
\label{Fig1}
	}
\end{figure}

For an interacting Fermi gas, the relevant quasiparticles are particle-hole excitations, where one particle is removed from the Fermi sea and a hole is created in its place.
If the Fermi gas is below the critical temperature for BCS superfluidity, this requires the breaking of a Cooper pair and the excitation has to overcome the pairing gap $\Delta$.
The second type of excitations in the system are collective excitations of the superfluid, which correspond to Bogoliubov-Anderson phonons and form the Goldstone mode of the system \cite{Nambu1960-ip,Goldstone1961-hs}. 

Experimentally, these physics can be studied using ultracold Fermi gases, where the strength of the interparticle interactions can be controlled via Feshbach resonances \cite{Chin2010-th}.
This makes it possible to adiabatically convert a BCS superfluid \cite{Bardeen1957-ti} of weakly bound Cooper pairs into a BEC of molecules \citep{Bloch2008-td,BCS-BEC_book2012}.
After the first observation of this BEC-BCS crossover in \cite{Leggett1980-sw, Nozieres1985-gq, Bartenstein2004-yz,Regal2004-rw}, various measurements of the change of the macroscopic properties of ultracold Fermi gases in the BEC-BCS crossover have been performed.
Starting from studies of collective oscillations \cite{Kinast2004-gs,Bartenstein2004-xh} experiments progressed to measurements of the speed of sound \cite{Joseph2007-jd} and critical velocity \cite{Miller2007-um,Weimer2015-gk}, and finally culminated in measurements of the equation of state \cite{Navon2010-bz,Nascimbene2010-xp,Horikoshi2010-sm,Ku2012-wd}.
Remarkably, the evolution of all these macroscopic quantities of the system can be linked to a single microscopic property of the system:
The size of the fermion pairs, which shrink from weakly bound Cooper pairs on the BCS side of the crossover to tightly bound molecules in the BEC regime.
The properties of these pairs have been explored by probing the excitation spectrum with techniques such as RF spectroscopy \cite{Chin2004-sm,Schunck2008-us,Schirotzek2008-ig,Stewart2008-aw}, fixed-momentum Bragg spectroscopy \cite{Hoinka2017-ej,Kuhn2020-ny} and RF dressing \cite{Behrle2018-uf}.
However, no measurement of the full low-energy excitation spectrum of fermionic superfluids in the BEC-BCS crossover has been performed.

In this work, we use momentum-resolved Bragg spectroscopy to measure the excitation spectrum of a homogeneous ultracold Fermi gas.
This allows us to directly observe the evolution of both single-particle excitations and collective modes in the BEC-BCS crossover. 
From our observations of the collective mode we extract the speed of sound in the system, while the shifting onset of the pair breaking continuum reveals the evolution of the superfluid gap throughout the BEC-BCS crossover.
Finally, we compare current state-of-the-art theories with our measurement of the gap.

For our experiments, we use an ultracold gas of $^6$Li atoms (Fig.\,\ref{Fig1}\,b) in a balanced spin mixture of the lowest two hyperfine states.
We follow an approach similar to the one taken in \cite{Gaunt2013-oj,Mukherjee2017-jd} and trap the gas in a cylindrical box potential whose walls are formed by blue-detuned laser beams.
This results in a system with an almost constant density per spin state of $n \approx 0.4/\mu\mathrm{m}^{-3}$, which corresponds to a Fermi energy of $E_F \approx h \times 7$\,kHz. 
The strength of the interparticle interactions is parametrized by the dimensionless parameter $1/k_Fa$, where $a$ is the s-wave scattering length and $k_F = (6 \pi^2 n)^{1/3}$ the Fermi wave vector.
The temperature of homogenous Fermi gases in the BEC-BCS crossover is challenging to measure \cite{Sobirey2020-gk}, but for systems with an interaction strength of $1/k_Fa = 0$ a technique based on measuring the total energy of the gas has been developed \cite{Yan2019-fh}.
For our system this approach gives us an estimate of $T/T_F \approx 0.13$, where $T$ is the temperature and $T_F = E_F/k_B$ is the Fermi temperature of the gas.  

\begin{figure*}[]
	\center{
	\includegraphics[width=18cm]{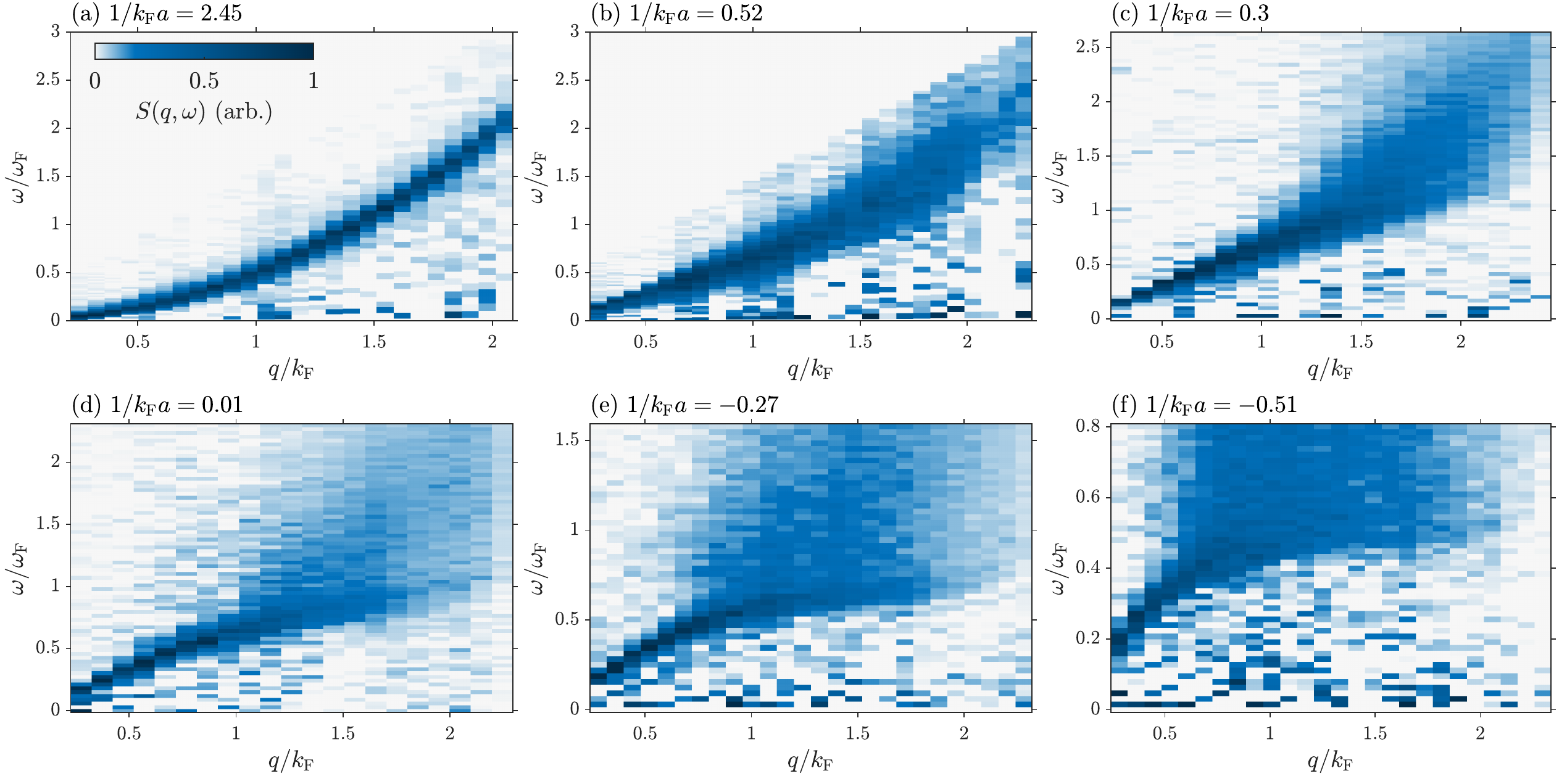}
	\caption{Evolution of the excitation spectrum in the BEC-BCS crossover.
	(a) In the deep BEC regime, the excitation spectrum follows the Bogoliubov dispersion of an interacting Bose gas, with a linear sound mode at low momenta and a quadratic dispersion of single-molecule excitations at high momenta. 
	(b,c) When moving into the crossover regime, the compressibility of the system decreases, and consequently the linear branch has a steeper slope and persists to higher momenta. 
	At the same time, the high-momentum part of the dispersion shows a strongly reduced curvature and starts to broaden, which indicates the transition to pair breaking excitations.
	(d) At the unitary point, there is already a strong pair breaking continuum, which becomes even more pronounced when going further into the BCS regime (e,f).
	\label{Fig2}}}
\end{figure*}

To measure the excitation spectrum of our system, we employ an experimental technique called Bragg spectroscopy ~\cite{Martin1988-fl,Stenger1999-bn,Lingham2016-si}.
This technique is based on applying two laser beams which are far detuned from the atomic transition so that single-photon scattering is strongly suppressed.
However, stimulated scattering processes, where a photon from one beam is scattered into the other, can occur if the difference in energy and momentum between the absorbed and emitted photon is transferred to the atoms (Fig.\,\ref{Fig1}\,a).
These two-photon scattering events therefore are only possible if the many-body system allows for the creation of excitations at this specific combination of transferred energy $\hbar \omega$ and momentum $\hbar q$.
By applying such Bragg beams and measuring the resulting heating rate $dE/dt$, we obtain the dynamic structure factor $S(q,\omega) \propto \omega^{-1}dE/dt$ \cite{Brunello2001-hk},  which quantifies the probability for an excitation with energy $\hbar\omega$ and momentum $\hbar q$ to be created and therefore describes the excitation spectrum of the system \cite{Supp}. 

For our first measurement, we prepare our gas at the so-called unitary point where the scattering length diverges and $1/k_Fa = 0$.
At this point, the only relevant length scale in the system is the inverse Fermi momentum $1/k_F$ and the system becomes scale invariant \cite{BCS-BEC_book2012,Ho2004-xl}. 
The gas is also very strongly interacting, with a collision rate that is comparable to the inverse Fermi time $E_F/h$ of the system.
This unitary Fermi gas is a canonical problem in many-body physics that was first posed in the context of neutron matter, and has come under intense experimental study with the development of ultracold Fermi gases.

Our measurement of the dynamic structure factor of the unitary Fermi gas is shown in Fig.\,\ref{Fig1}\,c.
The two distinct types of excitations discussed above are immediately visible. 
First, there is a narrow, well-defined mode whose energy is approximately proportional to its momentum, which we identify as the sound mode of the Fermi gas. 
For very low energies, where collisions have time to restore local thermal equilibrium, it can be understood in terms of hydrodynamics~\cite{Patel2020}, whereas for higher frequencies or weaker coupling strengths it is a Goldstone mode~\cite{Hoinka2017-ej,Kuhn2020-ny} that is driven by phase fluctuations of the superfluid order parameter. 

The second type of excitations are single particle excitations in which an atom is lifted out of the Fermi sea and a particle-hole excitation is created.
These particle-hole excitations appear as a broad continuum in our spectra, as each particle inside the Fermi sea can be excited to any unoccupied state if it receives the proper combination of energy and momentum transfer.
However, as the fermions are paired, this requires an energy of at least twice the pairing gap $\Delta$, resulting in a well-defined onset of the continuum.
The overall behavior of our measured dynamic structure factors is in excellent agreement with theoretical expectations \cite{Combescot2006-vk}; a comparison to a QRPA calculation of $S(q,\omega)$ is shown in the supplementary material \cite{Supp}.

While in the limits of small or large momentum transfer the response of the system can be clearly identified as either a collective or single particle excitation, there is a range of intermediate momenta where this is not as straightforward. 
In particular, as the collective mode approaches the pair breaking continuum it no longer follows the linear slope given by the speed of sound and instead starts to bend down.
This behavior is reminiscent of an avoided crossing with the onset of the pair breaking continuum, and indicates the existence of a coupling between the Goldstone mode and the excitation of single particles from the superfluid via pair breaking. 
Such a coupling has been predicted by theory \cite{Pieri2004-cq,Combescot2006-vk,Zou2016-rp,Hoinka2017-ej}, but had not yet been observed in experiments.

After examining the general structure of the excitation spectrum, we now proceed to measure the dynamic structure factor at interaction strengths ranging from the deep BEC to the BCS regime. 
The results are displayed in Fig.\,\ref{Fig2} and clearly show the evolution of the superfluid throughout the BEC-BCS crossover.

Our first observation is that the collective mode is present throughout the entire BEC-BCS crossover.
This is a direct consequence of the fact that the presence of a well-defined Goldstone mode is a fundamental feature of any neutral superfluid \cite{Anderson1958-ax,Hoinka2017-ej,Kuhn2020-ny}.
In  contrast, the nature of the single particle excitations changes completely when going across the crossover. 
On the BCS side of the resonance (Fig.\,\ref{Fig2}\,e,f), the pairs are large and weakly bound and we observe a broad continuum of pair breaking excitations. 
This continuum becomes less pronounced as the pairs become more tightly bound in the crossover regime and completely disappears from our spectra in the BEC regime (Fig.\,\ref{Fig2}\,a,b).
This is caused by the pairs turning into deeply bound molecules, which are only broken at very high energy and momentum transfers.
Consequently, when going towards the BEC regime pair-breaking is gradually replaced by a different single-particle excitation where a single unbroken molecule is ejected from the condensate.

This behavior directly shows the evolution of our system from a BCS superfluid of weakly bound Cooper pairs to a BEC of deeply bound molecules.
In the following, we discuss the properties of the collective mode and the pair breaking continuum in more detail and use them to extract quantitative information about our system

\begin{figure}[t]
	\center
	\includegraphics[width=8.6cm]{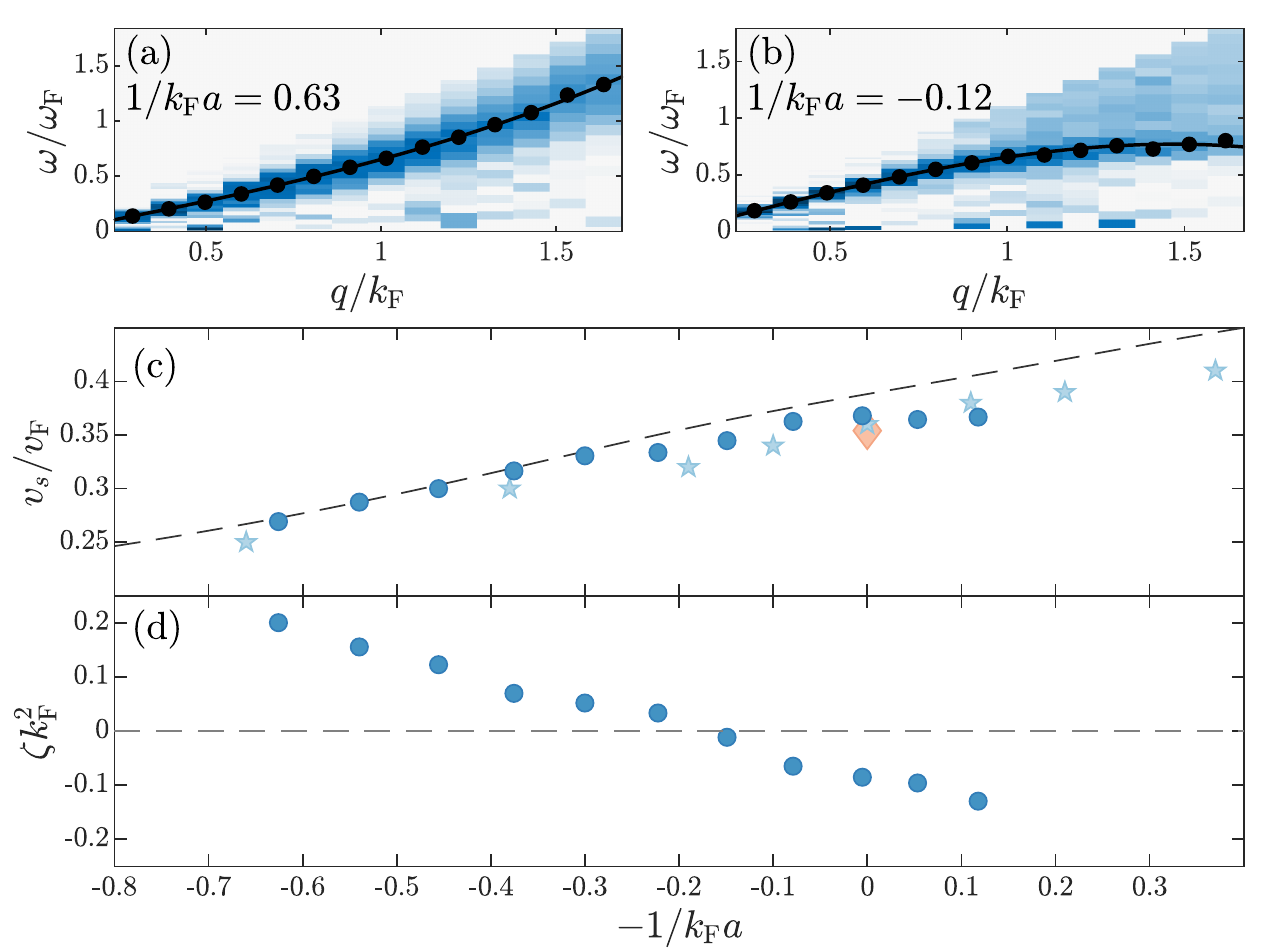}
	\caption{Measurements of the collective mode on the BEC (a) and BCS (b) side of the resonance.
	The black dots show the fitted center of the collective mode for each momentum slice, the black line is a fit according to the equation $\omega = v_s q (1 + \zeta q^2)$.
	(c) Speed of sound $v_s$ across the BEC-BCS crossover (blue dots) extracted from the fit to the collective mode.
	We find good agreement with a previous measurement of the speed of sound performed via fixed-momentum Bragg spectroscopy \cite{Hoinka2017-ej} (light blue stars), a measurement of the Bertsch parameter at unitarity \cite{Ku2012-wd} (orange diamond) and a quantum Monte Carlo calculation of the equation of state \cite{Astrakharchik2004-qs} (dashed line).
	(d) Prefactor $\zeta$ of the $q^3$ correction to the collective mode. 
	In the BEC regime, the dispersion bends upwards and $\zeta>0$.
	When moving towards the crossover regime, the value of $\zeta$ decreases until it changes sign at an interaction strength of $1/k_Fa \approx 0.2$.
	For interaction parameters $1/k_Fa \lesssim 0.2$, the collective mode bends down and $\zeta<0$.
	\label{Fig3}}
\end{figure}

First, we consider the behavior of the collective mode, whose curvature has important consequences for the damping processes allowed in the system and has been a topic of intense theoretical discussion.
\cite{Landau1949-az, Beliaev1958-vs}.
We follow \cite{Haussmann2009-mm,Kurkjian2016-uf,Zou2018-bu} and fit the dispersion with an expression of the form $\omega(q) = v_s q (1 + \zeta q^2)$, examples are shown in Fig.\,\ref{Fig3}\,(a,b). 
This captures both the change of the linear slope due to the changing speed of sound (Fig.\,\ref{Fig3}\,c) and the change in the curvature of the dispersion (Fig.\,\ref{Fig3}\,d).
We find that the dispersion is convex ($\zeta>0$) in the BEC regime, but when going towards the resonance $\zeta$ smoothly decreases until it changes sign at an interaction strength of $1/k_Fa\approx 0.2$ and the dispersion becomes concave ($\zeta<0$).
At unitarity, we obtain a value of $\zeta = -0.085(8)/k_F^2$, which is in good agreement with \cite{Haussmann2009-mm,Zou2018-bu} and provides a quantitative experimental benchmark. 

Next, we consider the properties of the pair breaking continuum. 
We find that the continuum shows a clear dependence on both the energy and momentum transfer (see e.g. Fig.\,\ref{Fig2}\,e).
On the energy axis, there is a sharp threshold of the continuum at a well defined energy, whereas the momentum axis shows a more gradual onset of pair breaking excitations. 
Both of these observations are directly related to important properties of the pairs.

The existence of an onset on the momentum axis can be understood by comparing the wavelength of the excitation to the size of the pairs.
If the size of the pairs is large compared to the wavelength of the excitation, a single particle can be excited and the pair can be broken.
However, if the pair is smaller than the wavelength of the Bragg lattice, the excitation exerts almost no differential force on the atoms and they are preferentially excited as an unbroken pair.
Therefore, as the size of the pairs changes in the BEC-BCS crossover, the onset of the continuum changes with the interaction strength.
In the BCS regime, the pairs are large and we observe a broad pair breaking continuum (\ref{Fig2}\,f).
Going through the crossover, the pairs become more tightly bound and the onset of the continuum correspondingly moves to higher momenta, until we reach the deep BEC regime of tightly bound molecules, where pair breaking excitations are strongly suppressed and no continuum is visible (\ref{Fig2}\,a).
In this regime, the gas has essentially become a strongly interacting Bose gas and pair breaking excitations only occur at very high momenta and energies.

\begin{figure}[t]
	\center
	\includegraphics[width=8.6cm]{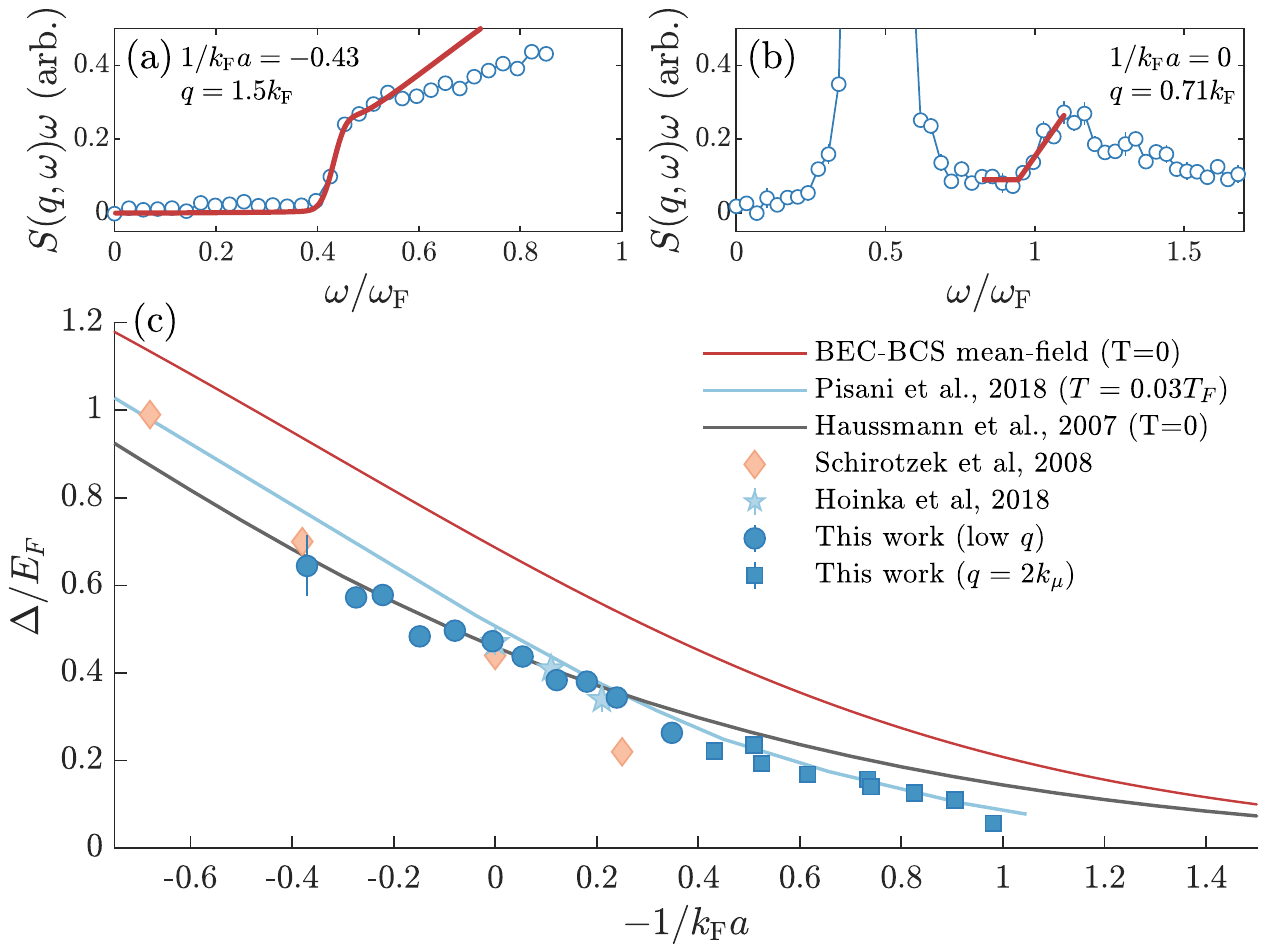}
	\caption{Measurement of the pairing gap in the BEC-BCS crossover.
	(a) Heating rate $S(q,\omega)\, \omega$ on the BCS side of the resonance ($1/k_Fa = -0.44$) measured at a fixed momentum transfer of $\hbar q = 1.5 \hbar k_F$.
	The onset of the pair breaking continuum is clearly visible in the data; the red line shows a fit to the onset that is used to extract the value of the pairing gap $\Delta$ (blue squares in panel (c)) \cite{Supp}.
	(b) Close to resonance, we perform measurements at low momentum to separate the onset of the pair breaking continuum from the collective mode, the results are shown in panel (c) as blue dots \cite{Supp}.
	(c) Pairing gap $\Delta$ across the BEC-BCS crossover.
	Our data is in good agreement with previous measurements (orange diamonds \cite{Schirotzek2008-ig}, blue stars \cite{Hoinka2017-ej}). 
	When comparing to theory, we find excellent agreement with self-consistent T-matrix calculations \cite{Haussmann2007-ph} close to resonance and in the BEC regime (black solid line), but towards the BCS regime calculations including Gor'kov-Melik-Barkhudarov corrections \cite{Pisani2018-sa} (orange line) are closer to our data.
	\label{Fig4}
	}
\end{figure}

The threshold on the energy axis is caused by the existence of the pairing gap $\Delta$, which describes the energy cost associated with breaking a Cooper pair.
We can therefore determine the evolution of the pairing gap by fitting the threshold of the pair breaking continuum in the dynamic structure factor, as shown in Fig.\,\ref{Fig4}\,a. 
This method works well in the BCS regime, but in the crossover the onset of the continuum is masked by the Goldstone mode (see Fig.\,\ref{Fig2}\,c,d).
In this regime, we therefore employ the method developed in ref. \cite{Hoinka2017-ej} and separate the pair breaking excitations from the Goldstone mode by strong driving at low momentum transfer.
An example of such a strongly driven spectrum can be seen in Fig.\,\ref{Fig4}\,b.

The gaps determined by our fits to the excitation spectra are shown in Fig.\,\ref{Fig4}\,c.
We find excellent agreement with previous experiments \cite{Hoinka2017-ej,Schirotzek2008-ig} that were performed in the BEC and crossover regimes.
Next, we compare our data to T-matrix calculations that self-consistently include strong pairing correlations (black line in Fig.\,\ref{Fig4}\,c \cite{Haussmann2007-ph}).
Taking the zero-temperature result, this theory is in excellent agreement with our data in the BEC and crossover regimes, but lies significantly above our measurements in the BCS regime.
While such a reduction of the gap could in principle be explained by finite temperature effects, the finite temperature results of the T-matrix calculation are inconsistent with our experimental observation that the system remains at almost constant entropy while ramping through the BEC-BCS crossover \cite{Supp}.
Another possible explanation could be that the size of the gap is influenced by particle-hole fluctuations.
These fluctuations are not expected to be important at unitarity, but lead to the famous Gor'kov-Melik-Barkhudarov (GMB) correction~\cite{Heiselberg2000,Gorkov1961-yf} in the BCS limit.
This effect is taken into account in a recent strong coupling calculation~\cite{Pisani2018-sa}, which is in good agreement with our data in the BCS regime, but lies significantly above our measurements on the BEC side of the resonance. 

In conclusion, we have presented momentum and energy resolved measurements of the excitation spectrum of a homogenous ultracold Fermi gas.
These measurements directly reveal the transformation from tightly bound molecules to weakly bound Cooper pairs that takes place in the BEC-BCS crossover.
Moreover, we have determined the evolution of both the slope and curvature of the Goldstone mode as well as the pairing gap in the BEC-BCS crossover, which provides quantitative benchmarks for theory.
These measurements are an excellent starting point for performing precision measurements of other key properties of strongly interacting Fermi gases, such as the critical temperature for superfluidity throughout the BEC-BCS crossover.
Our setup is also ideally suited to create imbalanced Fermi gases and study their excitation spectrum to search for exotic phases such as the elusive FFLO state \cite{Kinnunen2018-gb}.
Looking beyond our system, the combination of a homogeneous sample and momentum resolved Bragg spectroscopy established in this work is a powerful tool that can be used to measure the excitation spectrum of a wide variety of systems, ranging from dipolar gases to ultracold atoms trapped in optical lattices.

\begin{acknowledgments}
We thank R. Haussmann, L. Mathey, C. Vale, and W. Zwerger for helpful discussions and R. Haussmann, W. Zwerger and P. Pieri for providing us with the results of their calculations. 
This work is supported by the Deutsche Forschungsgemeinschaft (DFG, German Research Foundation) in the framework of SFB-925, Project No. 170620586, and the excellence cluster “Advanced Imaging of Matter,” EXC 2056, Project ID No. 390715994.
\end{acknowledgments}

\bibliography{bibliography}

\section{Supplementary Material}

\subsection{Preparation of homogeneous Fermi gases}
For our experiments, we use an ultracold gas of $^6$Li atoms in a balanced mixture of the lowest two hyperfine states, which we prepare as described in \cite{Weimer2015-gk}.
To obtain a homogeneous density distribution, we transfer this gas into a box potential formed by a combination of repulsive optical potentials.
The radial confinement is provided by a blue-detuned ($\lambda=532\,\mathrm{nm}$) ring potential that is projected onto the atoms using a high-resolution microscope. 
The optical setup used to create this ring potential uses a combination of three axicons and is described in detail in \cite{Hueck2018-sx}.
The vertical confinement is provided by two endcaps, which are formed by blue-detuned ($\lambda=532\,\mathrm{nm}$) elliptical laser beams intersecting the ring potential from the side.
As the diameter of the ring potential changes slightly over the vertical size of the potential, the shape of the box deviates from a perfect cylinder and takes the form of a truncated cone. 
Overall, the box has a height of $43\,\mu\mathrm{m}$ (FWHM) and an average radial extension of $50\,\mu\mathrm{m}$. 

\subsection{Density calibration}
To measure the density of our gas, we determine the two-dimensional column density $n_{\mathrm{2D}}$ with high intensity absorption imaging along the z-direction \citep{Hueck2017-cq}.  
By averaging over the central region where the vertical extent is not limited by the truncated cone and dividing it by the box height $b$, we obtain a first approximation of the three-dimensional density $\tilde{n} = n_{\mathrm{2D}}/b$.
However, this measurement can be affected by technical errors such as imperfect polarization and off-resonant light in the imaging beam, as well as systematic effects such as multiple scattering of photons in the optically dense sample.
Therefore, we calibrate the density by using a system with a known equation state as a reference.
We do this by preparing a unitary Fermi gas with similiar atom number in a hybrid trap, where the endcaps are left in place but the radial ring confinement is disabled.
Instead, the atoms are held in the radial direction by a weak magnetic trap that provides a harmonic confinement $V(r)=0.5 m \omega_r^2 r^2$  with $\omega_r = 2 \pi \cdot 29.8\;\mathrm{Hz}$. 
In this configuration, the local density $n(r)$ of the gas follows the harmonic confinement and decreases towards higher radii $r$ since the local chemical potential decreases as $\mu(r) = \mu(r=0) - V(r)$. 
In the central region of the cloud the local Fermi temperature $T_F(r)$ is high enough that $T/T_F(r) \ll 1$, and the local chemical potential is in good approximation given by $\mu(r) = \xi E_F(r)$, where $E_F(r)$ is the local Fermi energy and $\xi = 0.370(9)$ is the Bertsch parameter \cite{Ku2012-wd,Zurn2013-sk}.
This means that for a correctly calibrated density measurement a plot of $E_F(r)$ versus $V(r)$ must be linear with a slope of $-1/\xi$.
We can therefore obtain the corrected density $n = \alpha \tilde{n}$ of our system by introducing a correction factor $\alpha$ such that $dE_F(r)/dV(r)=-1/\xi$, with $E_F(r)= \hbar^2 (6 \pi \alpha \tilde{n}(r))^{2/3}/(2m)$.
We find a correction factor of $\alpha = 1.38(7)(5)$, where the first parenthesis gives our error in the determination of $dE_F(r)/dV(r)$ and the second parenthesis denotes the error due to the uncertainty of the Bertsch parameter.
This uncertainty in the density propagates to a relative systematic uncertainty in the Fermi energy of $4\%$ and $3\%$, respectively.

\subsection{Bragg spectroscopy}
We determine the excitation spectrum of our many-body system by using far-detuned laser beams to drive two-photon transitions and measuring the probability of creating excitations as a function of the transferred energy and momentum.
This technique is called optical Bragg spectroscopy and discussed in detail in \cite{Lingham2016-si}.
In our experiment, we implement this technique by using a high-resolution objective to project two intersecting $780$\,nm laser beams with waists of roughly $20\,\mu\mathrm{m}$ onto the atoms.
Two acousto-optic modulators set the frequency difference $\omega$ of the beams, while two motorized translation stages can be used to control the distance between the parallel beams at the entrance of the objective.
This in turn determines the crossing angle of the beams and thereby sets the momentum transfer $\vec{q}=\vec{k}_1-\vec{k}_2$ of the two-photon process, where $\vec{k}_1$ and $\vec{k}_2$ are the wavevectors of the intersecting beams. Since the system is isotropic, we restrict our discussion to the absolute value of the momentum transfer $q=|\vec{q}|$ along an arbitrary direction $\vec{e}_q = \vec{q}/q$.

\begin{figure}[t]
	\center
	\includegraphics[width=8.6cm]{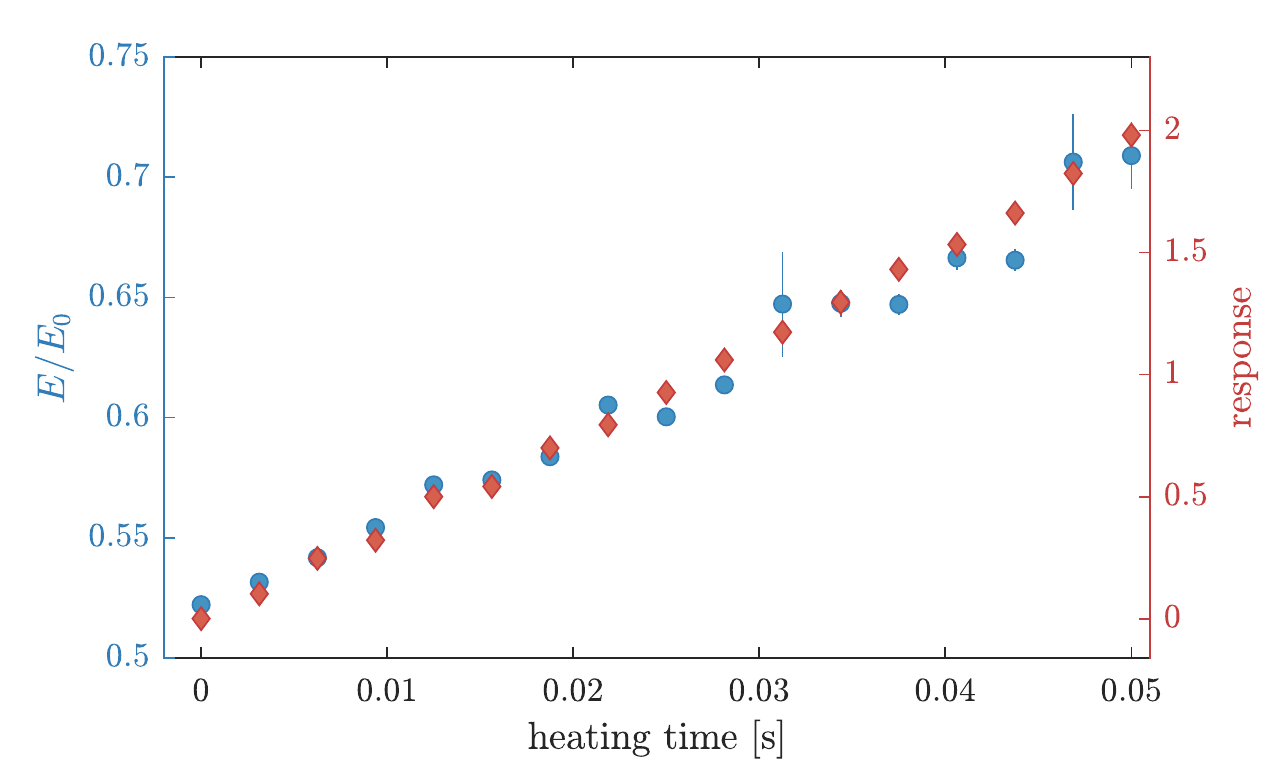}
	\caption{Energy measurement of a unitary Fermi gas in a box potential.
	The total energy, normalized by the energy of a non-interacting Fermi gas $E_0$ (blue dots), increases linearly with the heating time. 
	Here, the gas is transferred into a hybrid trap to determine the total energy from in situ images using the known equation of state. 
	The response (red diamonds) utilizing the condensate peak from matter wave focusing as a thermometer shows a similar linear behavior and can be used as an alternative to measure the heating rate.}
	\label{S1}
\end{figure}
 
The probability per unit time and particle to excite the many-body system from its ground state $|0\rangle$ by transferring the momentum $\hbar q$ and energy $\hbar \omega$ is given by Fermi's golden rule
\begin{equation*}
P(q,\omega) = 2\pi \Omega_R^2 S(q,\omega) 
\end{equation*}
with the dynamic structure factor
\begin{equation*}
S(q,\omega) =  \sum_n \left| \left\langle n \middle| \hat{\rho}^{\dagger}(q) \middle| 0 \right\rangle \right|^2 \delta(\omega - (E_n - E_0)/\hbar), 
\end{equation*}
the ground state energy $E_0$, excited states $|n\rangle$ with energy $E_n$, the Fourier transform of the density operator $\hat{\rho}(q) = \sum_{\vec{k}} \hat{a}_{\vec{k} + q\vec{e}_q}^{\dagger} \hat{a}_{\vec{k}}$ and the Rabi frequency $\Omega_R$ of the atom-light field coupling.
Consequently, the perturbation leads to a heating rate $dE/dt$ which is directly related to the dynamic structure factor $S(q,\omega)$ by
\begin{equation*}
\frac{dE}{dt} = \hbar \omega P(q,\omega) = 2 \pi \hbar \omega \Omega_R^2 S(q,\omega)
\end{equation*}
\cite{Brunello2001-hk} where we have neglected finite temperature effects as they are small for our superfluid system.

In our experiment, we have two different methods to measure the amount of energy that was deposited by the Bragg lattice.
On unitarity, we can follow a procedure established in \cite{Yan2019-fh} and directly measure the total energy of the system by releasing the gas into a harmonic potential and using the known equation of state \cite{Ku2012-wd} to calculate the total energy from the resulting density distribution. 
The results for Bragg pulses of different length are shown in Fig.\,\ref{S1}.
We observe that the energy increases linearly with the length of the Bragg pulse, which indicates that our measurements are performed in the linear response regime.
However, this method has a rather low signal-to-noise ratio and is quite sensitive to offsets in the density measurements. 
Therefore, we instead use the change of the condensate fraction in the BEC regime to determine the effect of the Bragg lattice on the system.

\begin{figure*}[tbp]
	\center
	\includegraphics[width=18cm]{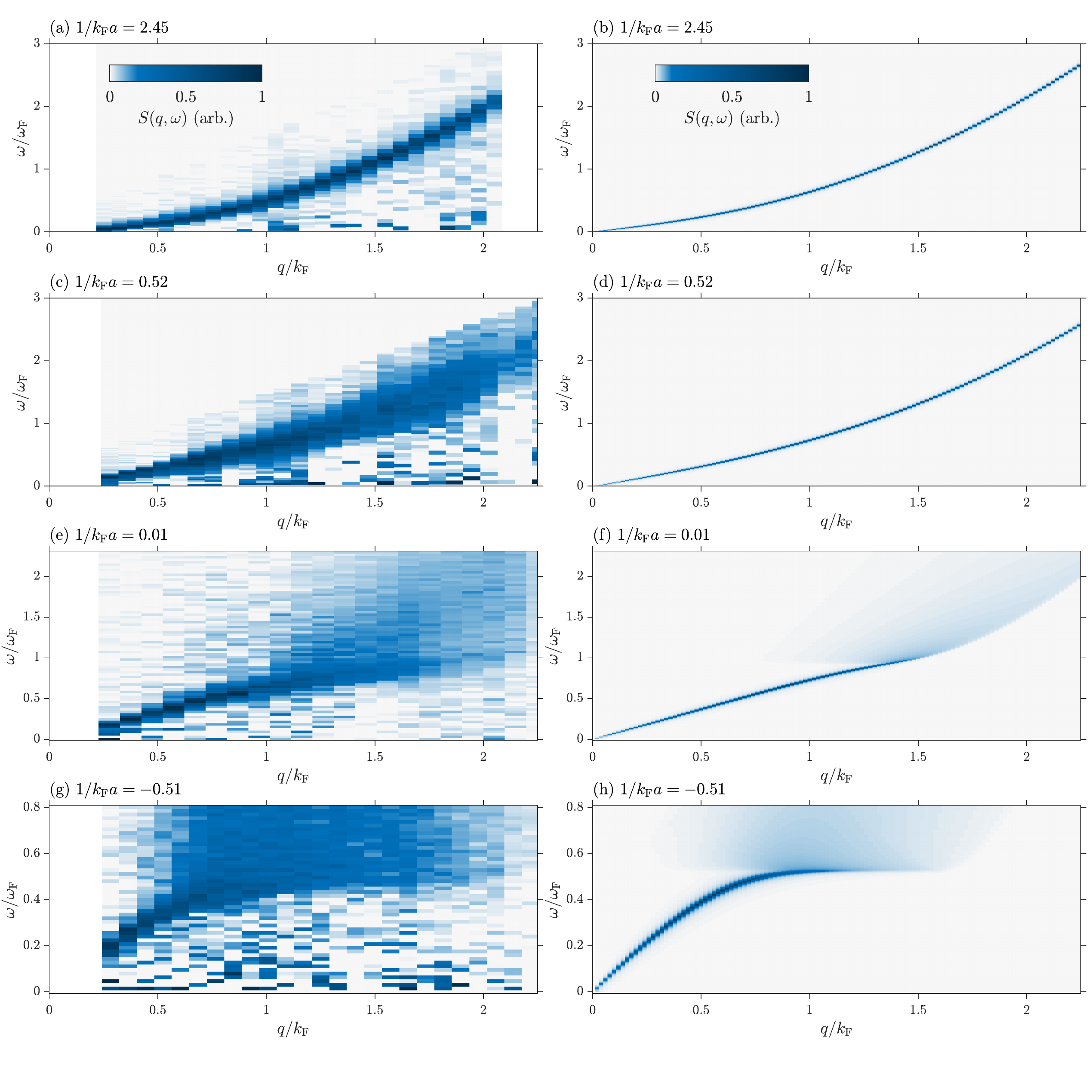}
	\caption{Comparison of the measured excitation spectra (left column) with spectra calculated using the quasiparticle random-phase approximation (right column). 
While the qualitative evolution of both the pair breaking continuum and the collective mode agrees, there are some notable differences.
On the BEC side, the collective mode is much narrower in the theoretical spectra. 
This can be a result of both finite temperature and instrumental broadening of the measurements, which could for example be caused by residual inhomogeneities of the gas.
As the spectra are normalized, the sharper collective mode in the theoretical spectra makes the continuum appear weaker, which is particularly notable for the unitary system.
However, despite this effect and the broadening present in the experimental data, important qualitative features such as the downbending of the collective mode and the overall shape of the continuum are still clearly visible in both the theoretical and experimental spectrum. 
In the BCS regime, there is excellent agreement between the experimental result and QRPA theory, with the only significant difference being the slightly different onset of the pair breaking continuum. 
This, however, is expected as the theoretical gaps used as inputs for the QRPA calculation are higher than our measurements in the BCS regime (see Fig.\,\ref{Fig4}).
Note that the increased noise at small energy transfers in the experimental spectra does not indicate the presence of excitations, but is an artifact of dividing a very small heating rate $dE/dt$ by a small frequency $\omega$ to obtain the dynamic structure factor $S(q,\omega)\propto 1/\omega \; dE/dt$. 
}
	\label{S4}
\end{figure*}

To do this, we define the response $r(q,\omega)=(A_0/A(q,\omega))-1$, where $A_0$ is the height of the condensate peak before and $A(q,\omega)$ is the height of the condensate peak after the application of the Bragg pulse \cite{Sobirey2020-gk}.
Similarly to the measurements of the total energy on unitarity shown in Fig.\,\ref{S1}, the response scales linearly with the length of the Bragg pulse (Fig.\,\ref{S1}).
This means that $r$ is proportional to total energy $\Delta E$ deposited by the Bragg pulse and can therefore be used to calculate the dynamic structure factor according to
\begin{equation*}
S(q,\omega) \propto \frac{r(q,\omega)}{\omega \; \Delta t}.
\end{equation*}

\subsection{Comparison of measured and calculated structure factors}

A comparison of the measured excitation spectra to state-of-the-art theoretical calculations is shown in Fig.\,\ref{S4}.
The calculation uses the quasiparticle random-phase approximation (QRPA), which is described in detail in ref. \cite{Hoinka2017-ej}, a very brief summary is given below.
QRPA calculations are performed by self-consistently solving the standard BCS-BEC mean-field equations while including an energy shift for the chemical potential to obtain the desired pairing gap.
For the data shown in Fig.\,\ref{S4} we use the results of the self-consistent T-matrix theory given in \cite{Haussmann2007-ph} as our input for the gap, as this theory provides a consistent data set over the whole BCS-BEC crossover region and is in reasonable agreement with experimental results (see Fig.\,\ref{Fig4}\,c).

Note that as we only have limited knowledge about the evolution of the temperature of the gas throughout crossover, we have chosen to compare our data to a QRPA calculation for an essentially zero-temperature system.
We have verified that this has only minor effects on the resulting spectra.

\subsection{Determination of the pairing gap}

According to ref. \cite{Combescot2006-vk} in systems with a positive chemical potential the onset of pair breaking is located at an energy transfer of $\omega = 2 \Delta$ for momenta $q<2 k_{\mu}=2 \sqrt{2 m \mu/\hbar^2}$.
As $\mu > 0$ for all interaction strengths covered in Fig.\,\ref{Fig4} we can therefore determine the size of the gap by finding the onset of the pair breaking continuum.
In the BCS regime, we perform Bragg spectroscopy at fixed momentum transfers of $q = 1.5 \ldots1.7\; k_F$ where the influence of the collective mode is negligible and fit the response $r(\omega)$ with a line-shape obtained using a QRPA calculation.
A sample fit is shown in Fig.\,\ref{Fig4}\,a.

As mentioned in the main text, this approach breaks down in the crossover regime as the onset of the continuum is masked by the Goldstone mode.
We therefore separate the pair breaking excitations from the Goldstone mode by strong driving at low momentum transfer.
As this strongly saturates the Goldstone mode, these spectra are not well described by the QRPA calculation which assumes the system to be in the linear response regime.
For these data points we therefore determine the onset of the pair breaking mode from the transition point of a phenomenological bilinear fit, an example of this is shown in Fig.\,\ref{Fig4}\,b.

\subsection{Temperature determination and adiabaticity of interaction ramps}

To obtain an estimate of the temperature of the system we again use the approach described in \cite{Yan2019-fh}.
It is based on performing an isoenergetic expansion of a Fermi gas on unitarity into a harmonic trap and using the known equation of state to determine the normalized total energy $E/E_0$ from the density distribution of the expanded system, where $E_0 = 3/5 N E_F$ is the total energy of a non-interacting Fermi gas.
We obtain a value of $E/E_0 = 0.43$, which according to the equation of state measured in \cite{Ku2012-wd} corresponds to an entropy per particle of $S/N k_B = 0.29$ and a temperature of $T/T_F = 0.128$, where $T_F=E_F/k_B$ is the Fermi temperature of the system and $k_B$ is the Boltzmann constant.
However, the determination of $E/E_0$ is sensitive to fringes and offsets on the density images. 
This is not a problem for systems close to or above the critical temperature, but becomes a serious issue for $T \lesssim 0.11 T_F$, where the energy of the gas becomes indistinguishable from the energy of a zero-temperature system.
For our energy measurement we obtain a statistical error of $E/E_0 = 0.43 \pm 0.02$, which corresponds to an error of $0.008\; T_F$ and $0.09\;N k_B$ for the temperature and entropy determination, respectively.
In addition, systematic effects in the density images can lead to additional errors.

\begin{figure}[t]
	\center
	\includegraphics[width=8.6cm]{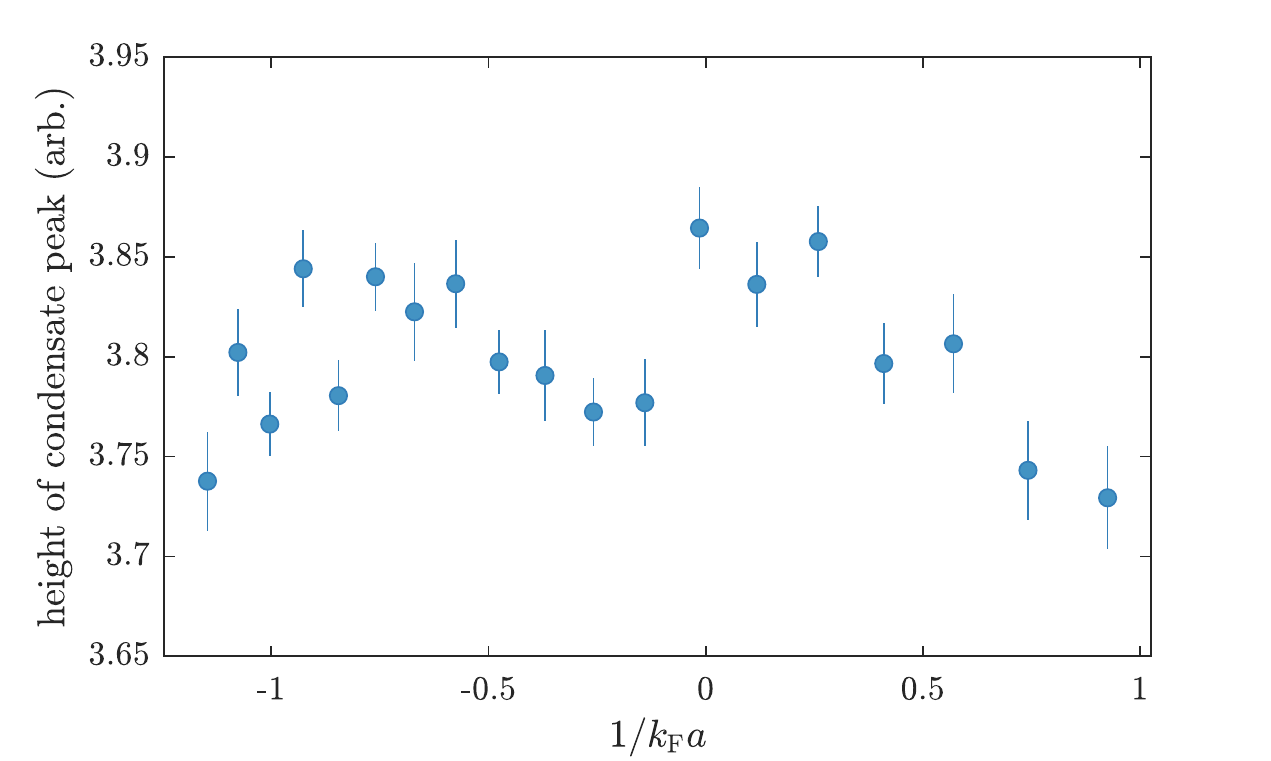}
	\caption{
Normalized height of the condensate peak after ramping to different interaction strengths in the BEC-BCS crossover.
The peak height varies by less than $4\%$.
Using the relationship between peak height and total energy found in Fig.\,\ref{S1}, this suggests a variation of the entropy per particle $S/N$ by about $0.07\,k_B$.
}
	\label{S2}
\end{figure}

Another limitation of this temperature determination is that it can only be used on unitarity, where the EOS is known with sufficient precision.
For our experiments, we prepare the system at a fixed interaction strength, perform our final stage of evaporative cooling, and then slowly ramp the interactions to the desired value.
While this does change the temperature of the gas, in a homogeneous system such as ours the entropy per particle remains unchanged when tuning the interactions in the system.
This makes $S/N$ a much more useful quantity for describing our system than the temperature $T/T_F$. 

To verify that our interaction ramps do not cause large amounts of technical heating we perform a set of measurements where we prepare our system as described above, ramp to different interaction strengths and hold the system for the same duration as in our gap measurements, and finally ramp to the BEC regime to measure the height of the condensate peak.
We find a variation in the normalized peak height of less than than $4\%$.
As the peak height scales with the entropy per particle this measurement allows us to estimate the variation of $S/N$ in our dataset.      
Using the relation between peak height and $E/E_0$ from Fig.\,\ref{S1} and assuming our value of $S/N = 0.29 k_B$ to be correct we estimate a variation in $S/N$ of less than $0.07 k_B$ for our measurements.

\begin{figure}[t]
	\center
	\includegraphics[width=8.6cm]{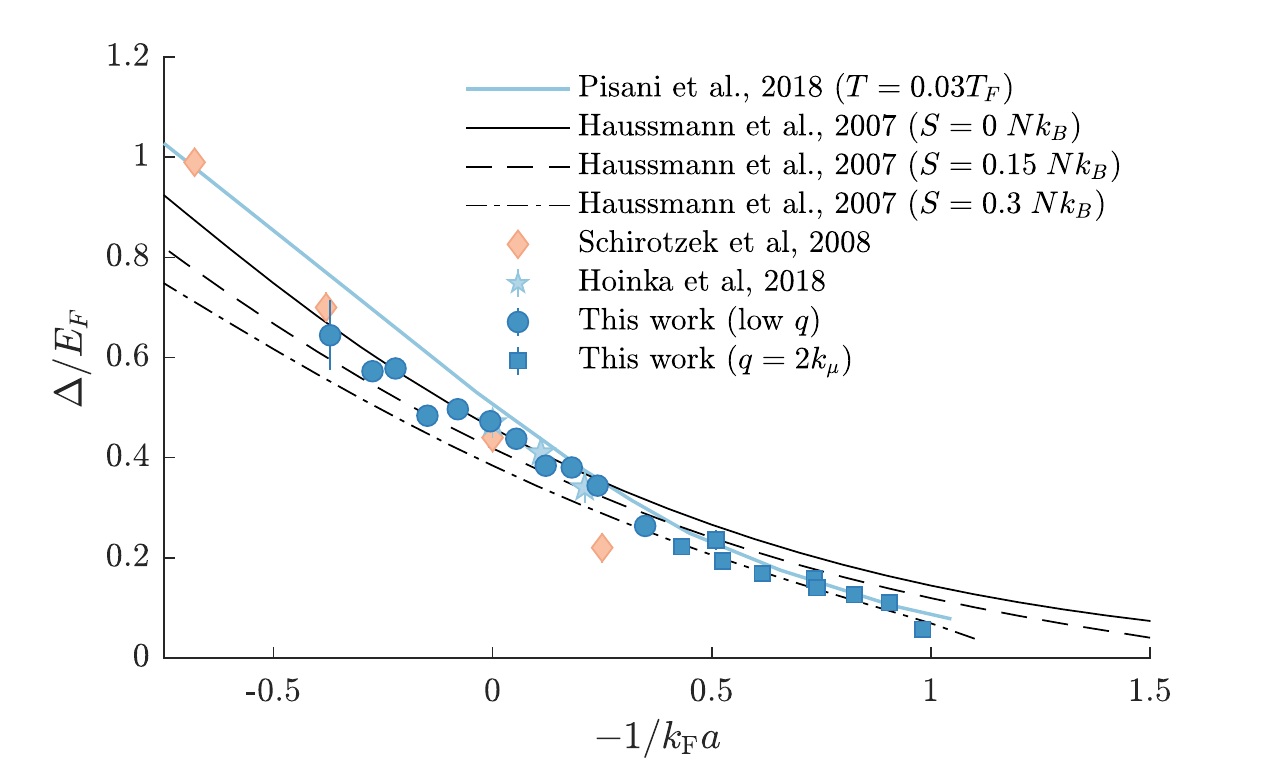}
	\caption{Measurement of the pairing gap in the BEC-BCS crossover in comparison to finite temperature T-matrix calculations. 
While the measured gap and zero-temperature theory are in excellent agreement in the crossover and BEC regimes, in the BCS regime calculations for a finite entropy per particle of $S/N=0.3 k_B$ are much closer the data. 
This is inconsistent with our observations, which suggest a nearly constant entropy of the gas throughout the BEC-BCS crossover.
}
	\label{S3}
\end{figure}

When comparing our data to finite-entropy theory (see Fig.\,\ref{S3}), we find that our data is compatible with an entropy per particle of $S/N \approx 0.1$\,$k_B$ in the crossover and BEC regimes.
However, in the BCS regime an entropy per particle on the order of $0.3$\,$k_B$ is required for data and theory to match.
This difference is significantly larger than the variation of $S/N$ of about 0.07\,$k_B$ we estimate for our experiments.

With all of these things considered, neither the T-matrix calculations from ref. \cite{Haussmann2007-ph} nor the approach from ref. \cite{Pisani2018-sa} are fully consistent with our data throughout the crossover.
The explanation for this could lie both on the theoretical and experimental side of the problem, and further work will be needed in both areas to resolve this question.

\begin{table*}
\parbox{.4\paperwidth}{
\begin{tabular}{ccll}
\hline
$B$ [G] & $\nicefrac{1}{k_F a}$ & $\nicefrac{v_s}{v_F}$ & $\zeta k_F^2$\\
\hline
800 & 0.63 & 0.269(5) & 0.2(1)\\
804 & 0.54 & 0.287(5) & 0.16(1)\\
808 & 0.46 & 0.3(5) & 0.12(1)\\
812 & 0.38 & 0.317(7) & 0.07(1)\\
816 & 0.3 & 0.331(7) & 0.05(1)\\
820 & 0.22 & 0.334(6) & 0.03(1)\\
824 & 0.15 & 0.345(8) & -0.01(1)\\
828 & 0.08 & 0.363(8) & -0.06(1)\\
832 & 0.01 & 0.368(6) & -0.085(8)\\
835 & -0.05 & 0.364(7) & -0.096(9)\\
839 & -0.12 & 0.367(7) & -0.129(8)\\
\hline
\end{tabular}
\caption{Fitted speed of sound $v_s$ and $q^3$-correction $\zeta$ of the collective mode shown in Fig. \ref{Fig3}. The errors give the statistical uncertainty of the fit.}
\label{TableCollectiveMode}
}
\hfill
\parbox{.4\paperwidth}{
\centering
\begin{tabular}{cclll}
\hline
$B$ [G] & $\nicefrac{1}{k_F a}$ & $\nicefrac{q}{k_F}$ & $\nicefrac{\Delta}{E_{\rm F}}$\\
\hline
812 &  0.37 & 0.69 & 0.64(7)\\
816 &  0.27 & 0.65 & 0.573(7)\\
820 &  0.22 & 0.7 & 0.58(1)\\
824 &  0.15 & 0.7 & 0.48(2)\\
828 &  0.08 & 0.71 & 0.497(7)\\
832 &  0.01 & 0.71 & 0.47(1)\\
835 & -0.05 & 0.6 & 0.44(2)\\
839 & -0.12 & 0.56 & 0.384(5)\\
843 & -0.18 & 0.49 & 0.38(1)\\
847 & -0.24 & 0.49 & 0.34(1)\\
855 & -0.35 & 0.28 & 0.26(1)\\
863 & -0.43 & 1.51 & 0.222(3)\\
867 & -0.51 & 1.57 & 0.24(2)\\
871 & -0.53 & 1.54 & 0.193(7)\\
879 & -0.62 & 1.57 & 0.1686(5)\\
886 & -0.73 & 1.62 & 0.16(1)\\
886 & -0.74 & 1.68 & 0.14(2)\\
894 & -0.83 & 1.7 & 0.126(2)\\
902 & -0.91 & 1.71 & 0.11(1)\\
910 & -0.98 & 1.72 & 0.057(7)\\
\hline
\end{tabular}
\caption{Fitted values for the pairing gap shown in Fig. \ref{Fig4}. In addition, the respective momentum transfer $q$ for each measurement of the continuum threshold is shown. The errors give the statistical uncertainty of the fit.}
\label{TableGap}
}
\end{table*}

\end{document}